\documentclass[a4paper,12pt,twoside, final]{article}
\usepackage[dvips]{graphics}
\usepackage{epsf}
\usepackage{graphicx}
\usepackage{epsfig}
\usepackage{amsmath}
\usepackage[psamsfonts]{amssymb}
\usepackage[psamsfonts]{eucal}
\usepackage{amsthm}
\usepackage[utf8]{inputenc}
\usepackage{textcomp}
\usepackage{mathptmx}
\usepackage[scaled]{helvet}
\usepackage{float}

\swapnumbers
\theoremstyle{plain}
\theoremstyle{definition}
\title{Relativity, the Special Theory,\\
 explained to Children\\
(from 7 to 107 years old)}

\author{Charles-Michel Marle\\
  Professor retired from Institut de Math\'ematiques de Jussieu\\
  and from Universit{\'e} Pierre et Marie Curie (today Sorbonne Université)\\
  Paris,  France}
\date{}
\begin{document}
\maketitle

\begin{abstract}
The author thinks that the main ideas or Relativity Theory can  be
explained to children (around the age of 15 or 16) without
complicated calculations, by using very simple arguments of affine
geometry. The proposed approach is presented as a conversation
between the author and one of his grand-children. Limited here to
the Special Theory, it will be extended to the General Theory
elsewhere, as sketched in conclusion.

\end{abstract}

%
\centerline{For Agathe, Florent, Basile, Mathis, Gabrielle,}
\centerline{Morgane, Quitterie, Marilou} 
\centerline{and my future other grand- and grand-grand-children }

\section{Prologue}
Maybe one day, one of my grand-children, at the age of 15
or 16, will ask me:
\par\smallskip

{---}\quad Grand-father, could you explain what is Relativity
Theory? My Physics teacher lectured about it, talking of rolling
trains and of lightnings hitting the railroad, and I understood
almost nothing!
\par\smallskip

This is the discussion I would like to have with her (or
him).
\par\smallskip

{---}\quad Do you know the theorem: the diagonals of a
parallelogram meet at their middle point?
\par\smallskip

{---}\quad Yes, I do! I even know that the converse is true: if
the diagonals of a plane quadrilateral meet at their middle point,
that quadrilateral is a parallelogram. And I believe that I know a
proof!
\par\smallskip

{---}\quad Good! You know all the stuff needed to understand
the basic idea of Relativity theory! However, we must first think
about Time and Space.
\par\smallskip

{---}\quad Time and space seem to me very intuitive, and yet
difficult to understand in deep!
\par\smallskip

{---}\quad  Many people feel the same. The true nature of Time and
Space is mysterious. Let us say that together, Time and Space make
the frame in which all physical phenomena take place, in which all
material objects evolve, including our bodies. We should keep a
modest mind profile on such a subject. We cannot hope to understand
all the mysteries of Time and Space. We should only try to
understand some of their properties and to use them to describe
physical phenomena. We should be ready to change the way we think
about Time and Space, if some experimental evidence shows that we
were wrong.
\par\smallskip

{---}\quad But if we do not know what are Time and Space, how
can we hope to understand some of their properties, and to be able
to use them?
\par\smallskip

{---}\quad By building mental pictures of Time and Space.
Unfortunately we, poor limited human beings, cannot do better: we
know the surrounding world only through our senses (enhanced by the
measurement and observation instruments we have built) and our
ability of reasoning. Our reasoning always apply to the mental
pictures we have built of reality, not to reality itself.
\par\smallskip

Let me now indicate how the mental pictures of Time and Space
used by scientists have evolved, mainly from Newton to Einstein.

\section{The views of Newton and Leibniz about Time and Space}

\subsection{Newtonian Time\hfill}
\par\smallskip

The great scientist Isaac Newton \cite{newton} (1642--1727) used, as
mental picture of Time, a straight line $\cal T$, going to infinity
on both sides, hence with no beginning nor end and  no privileged
origin. Each particular time, for example \lq\lq now\rq\rq, or
\rq\rq three days ago at the sunset at Paris\rq\rq, corresponds to a
particular element of that straight line.
\par\smallskip

Observe that Newton considered, without any discussion, that
for each event happening in the universe, there was a corresponding
well defined time (element of the straight line $\cal T$), the time
at which that event happens.
\par\smallskip

{---}\quad Where is that straight line $\cal T$? Is it drawn
in some plane or in space?
\par\smallskip

{---}\quad Nowhere! You should not think about the straight
line of Time $\cal T$ as drawn in something of larger dimension.
Newton considered Time as an abstract straight line, because
successive events are linearly ordered, like points on a straight
line. Don't forget that $\cal T$ is a mental picture of Time, not
Time itself! However, that mental picture is much more than a
confuse idea: it has very well defined mathematical properties. In
modern language, we say that $\cal T$ is endowed with an
\emph{affine structure} and with an \emph{orientation}.
\par\smallskip

{---}\quad What is an affine structure? and what is its use?
\par\smallskip

{---}\quad An affine structure on a line allows us to compare two time
intervals and to take their ratio, for example to say that one of
these intervals is two times the other one.
Newton considered the comparison of two time intervals as
possible, even when they were many centuries or millenaries apart,
and to take their ratio. In modern mathematical language, that
property determines on a line an \emph{affine structure}.
\par\smallskip
For the mathematician, that property means that we can apply
transforms to $\cal T$ by sliding it along itself, without
contraction nor dilation, and that these transforms (called
\emph{translations}) do not change its properties.
\par\smallskip

For the physicist, it means that the physical laws which govern the evolution 
with time of any system remain the same at all times.
\par\smallskip

Another important property of Time: it always flows from past to
future. To take it into account, we endow $\cal T$ with an
\emph{orientation}; it means that we consider the two directions (from
past to future and from future to past) as different, not
equivalent, for example by choosing the direction from past to
future as preferred. We then say that $\cal T$ is \emph{oriented}.

\subsection{Newton's absolute space}

{---}\quad OK, I roughly agree with that mental picture,
although it does not account for the main property of Time: it flows
continuously and we cannot stop it! And what about Space?
\par\smallskip

{---}\quad Newton identified Space with the three dimensional space
of geometers, denoted by $\cal E$~: the space in which there are
various figures made of planes, straight lines, spheres, polyhedra,
which obey the theorems developed in Euclidean geometry: Thales and
Py\-tha\-go\-ras theorems, the theorem which says that the diagonals of a
parallelogram meet at their middle point, $\ldots$

\subsection{The concept of Space-Time}

Newton used Time and Space to describe the motion of every object
$A$ of the physical world as follows. That object occupies, at each
time $t$ (element of $\cal T$) for which it exists, a position $A_t$
in Space $\cal E$. The motion of of $A$ is described by its
successive positions $A_t$ when $t$ varies in $\cal T$.
\par\smallskip

Let me introduce now a new concept, that of Space-Time
\cite{principles}, due to the German mathematician Hermann Minkowski
(1864--1909). That concept was not used in Mechanics before the
discovery of Special Relativity. That is very unfortunate, since its
use makes much easier the understanding of the foundations of
Classical Mechanics, as well as those of Relativistic Mechanics.
Therefore I use it now, with the absolute Time and Space of Newton,
although Newton himself did not use that concept.
\par\smallskip

Newton Space-Time is simply the product set ${\cal E}\times{\cal
T}$, whose elements are pairs (called \emph{events}) $(x,t)$, made
by a point $x$ of $\cal E$ and a time $t$ of $\cal T$.
\par\smallskip

{---}\quad What is the use of that Space-Time?
\par\smallskip

{---}\quad It is very convenient to describe motions. For example,
the motion of a material particle $a$ (a very small object whose
position, at each time $t\in{\cal T}$, is considered as a point
$a_t\in{\cal E}$), is described by a line in ${\cal E}\times{\cal
T}$, made by the events $(a_t,t)$, for all $t$ in the interval of
time during which $a$ exists. That line is called the \emph{world
line} of $a$.
\par\smallskip
You will see on Figure~1 (where, for simplicity, Space is
represented as a straight line, as if it were one-dimensional) the
world lines of three particles, $a$, $b$ and $c$.

\begin{figure}
  $$\scalebox{1.2}{\includegraphics[
  ]{relpic.1}}$$
  \caption{World lines in Newton Space-Time.}
\end{figure}

\begin{itemize}

\item The world line  of $b$ est parallel to the Time axis $\cal T$:
that particle is at rest, il occupies a fixed position in the
absolute Space $\cal E$.

\item The world line of $c$ is a slanting straight line. The
trajectory of that particle in absolute Space  $\cal E$ is a
straight line and its velocity is constant.

\item The world line of $a$ is a curve, not a straight line. It means
that the velocity of $a$ changes with time.

\end{itemize}

\subsection{Absolute rest and motion}

For Newton, \emph{rest} and \emph{motion}  were
absolute concepts: a physical object is at rest if its position in
Space does not change with time; otherwise, it is in motion.
\par\smallskip

{---}\quad It seems very natural. Why should we change this view?
\par\smallskip

{---}\quad Because nothing is at rest in the Universe! The
Earth rotates around its axis and around the Sun, which rotates
around the center of our Galaxy. And there are billions of galaxies
in the Universe, each of them moving with respect to the others! For these
reasons, Newton's concept of an absolute Space was criticized very
early, notably by his contemporary, the great mathematician and
philosopher Gottfried Wilhelm Leibniz (1647--1716).

\subsection{Reference frames}

{---}\quad But without knowing what is at rest in the
Universe, how Newton managed to study the motions of the planets?
\par\smallskip

{---}\quad To study the motion of a body $A$, Newton, and after him
almost all scientists up to now, used a \emph{reference frame}. It
means that he used another body $R$ which remained approximately
rigid during the motion he wanted to study, and he made as if that
body was at rest. Then he could study the \emph{relative motion} of
$A$ with respect to $R$.
 \par\smallskip

Assuming that Newton's absolute Space $\cal E$ exists, we
recover the description of absolute motion of $A$ by choosing, for
$R$, a body at rest in $\cal E$. The corresponding reference frame
is called the \emph{absolute fixed frame}.
\par\smallskip

The body $R$ used to determine a reference frame can be, for
example,
\begin{itemize}

\item  the Earth (if we want to study the motion of a falling apple),

\item the trihedron made by the straight lines which join the center
of the Sun to three distant stars (if we want to study the motions
of the planets in the solar system).

\end{itemize}

\subsection{Galilean frames and Leibniz Space-Time}

All reference frames are not equivalent. A \emph{Galilean
frame}~\footnote{\ In memory of Galileo Galilei, (1564--1642), the
founder of modern Physics.}, also called an
\emph{inertial frame}, is a reference frame in which the
\emph{principe of inertia} holds true. That principle, first formulated
for absolute motions in Newton's absolute space $\cal E$,
says that the (absolute) motion of a free particle takes place on a
straight line, at a constant speed. But, as shown by Newton himself,
that principe remains true for the \emph{relative motion} of a free
particle with respect to some particular reference frames, the
\emph{Galilean frames}.
\par\smallskip

More exactly, let us assume that the principle of inertia holds true
for the relative motion of free particles with respect to the
reference frame defined by the rigid body $R_1$. What happens for
the relative motion of these free particles with respect to another
reference frame, defined by another rigid body $R_2$? It is easy to
see that the principle of inertia still holds true \emph{if and only
if} the relative motion of $R_2$ with respect to $R_1$ is a motion
by translation at a constant speed.
\par\smallskip

The absolute frame, if it exists, therefore appears as a Galilean
frame among an infinite number of other Gallilean frames, that no
measurement founded on mechanical properties can distinguish from
the others. For this reason, several scientists, following Leibniz,
doubted about its existence.
\par\smallskip

Leibniz accepted Newton's concept of an absolute Time, but not that
of an absolute Space. His views were not successful during his life,
probably because at that time nobody saw how to cast them in a
mathematically rigorous setting. Now we can do that; let me explain
how.
\par\smallskip

We will consider that at each time $t\in{\cal T}$, there exists a
\emph{Space at time $t$}, denoted by ${\cal E}_t$, whose properties
are those of the three-dimensional Euclidean space of geometers. We
must consider that the Spaces ${\cal E}_{t_1}$ and ${\cal E}_{t_2}$,
at two different times $t_1$ and $t_2$, $t_1\neq t_2$, have no
common element. Leibniz Space-Time, which will be denoted by $\cal
U$ (for Universe), is the disjoint union of all the Spaces ${\cal
E}_t$ for all times $t\in{\cal T}$. So, according to Leibniz views,
we still have a Space-Time, but no more an absolute space~! The next
picture shows,
\begin{itemize}
\item on the left side, Newton Space-Time ${\cal E}\times{\cal T}$, with
the two projections $p_1:{\cal E}\times{\cal T}\to{\cal E}$ and
 $p_2:{\cal E}\times{\cal T}\to{\cal T}$;

\item on the right side, Leibniz Space-Time $\cal U$, endowed with only
one natural projection onto absolute Time $\cal T$, still denoted by
$p_2:{\cal U}\to{\cal T}$; the horizontal lines represent the Spaces
${\cal E}_t=p_2^{-1}(t)$, for various values of $t\in{\cal T}$.
\end{itemize}

\begin{figure}
  $$\scalebox{1.2}{\includegraphics[
  ]{relpic.2}}$$
  \caption{Newton and Leibniz Space-Time.}
\end{figure}

{---}\quad But how do you put together the Spaces at various times
${\cal E}_t$ to make Leibniz Space-Time $\cal U$? Are they stacked
in an arbitrary way?
\par\smallskip

{---}\quad Of course no! Leibniz Space-Time $\cal U$ is a
$4$-dimensional affine space, fibered ({\it via} an affine map) over
Time $\cal T$, which is itself a $1$-dimensional affine space. Its
fibres, the Spaces ${\cal E}_t$ at various times $t\in{\cal T}$, are
$3$-dimensional Euclidean spaces. The affine structure of $\cal U$
is determined by the \emph{principle of inertia} of which we have
already spoken. That principle can be formulated in a way which does
not use reference frames, by saying:
 \par\smallskip

\centerline{\emph{The world line of any free particle is a straight
line.}}
\par\smallskip

So formulated, the principle of inertia can be applied to Newton
Space-Time ${\cal E}\times{\cal T}$ and to  Leibniz Space-Time $\cal
U$ as well. More, it \emph{determines} the affine structure of $\cal
U$, since one can easily show that the affine structure for which it
holds true, if any, is unique. A physical law, the \emph{principle
of inertia}, is so embedded in the geometry of Leibniz Space-Time
$\cal U$.
\par\smallskip

By using a reference frame $R$, one can split Leibniz Space-Time
into a product of two factors: a space ${\cal E}_R$,  fixed with
respect to that frame, and the absolute Time $\cal T$. But of
course, the space ${\cal E}_R$ depends on the choice of the
reference frame $R$. For that reason, it seems that before 1905, not
many scientists were aware of the fact that by dropping Newton's
absolute Space $\cal E$, they already had completely changed the
conceptual setting in which motions are described:

\begin{itemize}

\item according to Newton, absolute Space ${\cal E}$ and absolute Time
${\cal T}$ were directly related to reality, while Space-Time ${\cal
E}\times{\cal T}$ was no more than a mathematical object, not very
interesting (he did not use it) and not directly related to reality;

\item but according to Leibniz's views, when expressed as done above,
it is Space-Time $\cal U$ which is directly related to reality, as
well as absolute Time $\cal T$; absolute Space $\cal E$ no more
exists.

\end{itemize}
\section{Relativity}

Einstein \cite{principles} was led to drop Leibniz Space-Time when
trying to reconcile the theories used in two different parts of
Physics: Mechanics on one hand, Electromagnetism and Optics on the
other hand.
\par\smallskip

According to the theory built by the great Scotch physicist James
Clerk Maxwell (1831--1879),  electromagnetic phenomena propagate in
vacuum as waves, with the same velocity in all directions,
independently of the motion of the source of these phenomena.
Maxwell soon understood that light was an elecromagnetic wave, and
lots of experimental results confirmed his views.

\subsection{The luminiferous ether, a short lived hypothesis}

In Leibniz Space-Time (as well as in Newton Space-Time)
\emph{relative velocities behave additively}. In that setting, it is
with respect to \emph{at most one particular reference frame} that
light can propagate with the same velocity in all directions.
Physicists introduced a new hypothesis: electromagnetic waves were
considered as vibrations of an hypothetic, very subtle, but highly
rigid medium called the \emph{luminiferous ether}, everywhere
present in space, even inside solid bodies. They thought that it was
with respect to the ether's reference frame that light propagates at
the same velocity in all directions. This new hypothesis amounts to
come back to Newton's absolute Space identified with the ether.
There were even physicists who introduced additional complications,
by assuming that the ether, partially drawn by the motion of moving
bodies, could deform with time!
\par\smallskip

{---}\quad But if the luminiferous ether really exists, accurate
measurements of the velocity of light in all directions should allow
the determination of the Earth's relative velocity with respect to
the ether!
\par\smallskip

{---}\quad Good remark! These measurements were made several times,
notably by Albert Abraham Michelson (1852--1931) and Edward Williams
Morley (1838--1923), between 1880 et 1887.  No relative velocity of
the Earth with respect to the luminiferous ether could be detected.
\par\smallskip

These results remained not understood until 1905, despite many
attempts. The most interesting of these attempts was that due to
Hendrik Anton Lorentz (1853--1928) and George Francis FitzGerald
(1851--1901). Independently, they proposed the following hypothesis:
when a rigid body, for example a rule or the arm of an
interferometer, is moving with respect to the luminiferous ether,
that body contracts slightly in the direction of its relative
displacement.
\par\smallskip

{---}\quad So that is the famous relativistic contraction my teacher
spoke about!
\par\smallskip

{---}\quad No! Not at all! Lorentz and FitzGerald considered that
contraction as a true physical effect of the relative motion of a
body with respect to the ether.
This assumption is now completely abandoned, together with the
luminiferous ether! The relativistic contraction of lengths and
dilation of times has nothing to do with
it: rather than a real phenomenon, it is only an appearance, like
the following effect of perspective. Imagine that you look at a 20
centimeters rule, from a distance of, say two meters from its
center. That rule looks shorter when it is not perpendicular to the
straight line which joins your eye to its center than when it is. It
may even seem to be reduced to a point when it lies along that
straight line. As we will soon see, the relativistic contraction of
lengths and dilation of times has a similar origin.

\subsection{Minkowski Space-Time}

Einstein was the first~\footnote{\ The great French mathematician
Jules Henri Poincar\'{e} (1854--1912) has, almost simultaneously and
independently, presented very similar ideas \cite{poinca}, without
explicitly recommending to drop the concept of an absolute Time.} to
understand (in 1905) that the results of Michelson and Morley
experiments could be explained by a deep change of the properties
ascribed to Space and Time. At that time, his idea appeared as truly
revolutionary. But now it may appear as rather natural, if we think
along the following lines:
\par\smallskip

\emph{When we dropped Newton Space-Time in favour of Leibniz
Space-Time, we recognized that there is no absolute Space,  but that
Space depends on the choice of a reference frame. Maybe Time too is
no more absolute than Space, and depends on the choice of a
reference frame!}
\par\smallskip

{---}\quad But if we drop absolute Time, which properties are left
to our Space-Time?
\par\smallskip

{---}\quad In 1905, Einstein implicitly considered that Space-Time
still was a $4$-dimensional affine space, which will be called
\emph{Minkowski Space-Time} and will be denoted by $\cal M$. He
implicitly considered too that \emph{translations} of $\cal M$ leave
its properties unchanged, and he assumed that the \emph{principe of
inertia} still holds true in $\cal M$ when expressed without the use
of reference frames:

\centerline{\emph{The world line of any free particle is a straight
line.}}
\par\smallskip

He also kept the notion of a \emph{Galilean frame}. In $\cal M$, a
Galilean frame is  determined by a direction of straight line (not
any straight line, a \emph{time-like} straight line, as we will see
below). Given a Galilean frame $R$, Minkowski Space-Time $\cal M$
can be split into a product ${\cal E}_R\times{\cal T}_R$ of a
three-dimensional Space ${\cal E}_R$ and a one-dimensional Time
${\cal T}_R$, which both depend on $R$. Let me recall that in
Leibniz Space-Time $\cal U$, a Galilean frame $R$ allowed us to
split $\cal U$ into a product ${\cal E}_R\times{\cal T}$ of a
three-dimensional Space ${\cal E}_R$, which depended on $R$, and the
one-dimensional absolute Time $\cal T$, which did not depend on $R$.
That is the main difference between Leibniz's and Einstein's views
about Space and Time.
\par\smallskip

Under these hypotheses, the properties of Space-Time follow from two
principles:

\begin{itemize}

\item the \emph{Principle of Relativity:} all physical laws have the
same expression in all Galilean frames;

\item the \emph{Principle of Constancy of the velocity of light:} the
modulus of the velocity of light is an universal constant, which
depends neither on the Galilean frame with respect to which it is
calculated, nor on the motion of the source of that light.
\end{itemize}

{---}\quad You said that a direction of straight line was enough to
determine a Galilean frame. But how is that possible, since we no
more have an absolute Time?
\par\smallskip

{---}\quad That determination will follow from the pinciple of
constancy of the velocity of light. Let us call \emph{light lines}
the straight lines in $\cal M$ which are possible world lines of
light signals. Given an event $A\in{\cal  M}$, the light lines
through $A$ make a $3$-dimensional cone, the \emph{light cone with
apex $A$}; the two layers of that cone are called \emph{the past
half-cone} and \emph{the future half-cone} with apex $A$. Since it
is assumed that translations leave unchanged the properties of
Space-Time, the light cone with another event $B$ as apex is deduced
from the light cone with apex $A$ by the translation which maps $A$
onto $B$.
\par\smallskip

Apart from light lines, there are two other kinds of straight lines
in $\cal M$:
\begin{itemize}

\item \emph{time-like straight lines}, which lie \emph{inside} the
light cone with any one of their elements as apex;

\item and \emph{space-like straight lines}, which lie \emph{outside}
the light cone with any of their element as apex.
\end{itemize}

\begin{figure}[H]
  $$\scalebox{1.2}{\includegraphics[
  ]{relpic.3}}$$
  \caption{Construction of  Space and Time relative to a Galilean frame.}
\end{figure}

I can now explain how the direction of a time-like straight line
$\cal A$ determines a Galilean frame $R$. That frame is such that
the rigid bodies at rest in it are those whose all material points
have, as world lines, straight lines parallel to $\cal A$. The
straight lines parallel to $\cal A$ will be called the
\emph{isochorous lines}~\footnote{\ The word \emph{isochorous},
already used in Thermodynamics, refers here to a set of events which
all occur at the same spatial location at various times, in
similarity with the word \emph{isochronous} which refers to a set of
events which all occur simultaneously in time at various spatial
locations.} of the reference frame $R$; each of these lines is a set
of events which all happen at the same place in the Space ${\cal
E}_R$ of our frame $R$. For each event $M\in{\cal M}$, the set of
all other events which occur at the same time as $M$, for the Time
${\cal T}_R$ of our Galilean frame $R$, will be called the
\emph{isochronous subspace} through $M$, for the Galilean frame $R$.
It is a $3$-dimensional affine subspace ${\cal E}_{R,\,M}$ of $\cal
M$ containing the event $M$, and the other isochronous subspaces for
$R$ are all the $3$-dimensional subspaces of $\cal M$ parallel to
${\cal E}_{R,M}$. They are determined by the property: the length
covered by a light signal, calculated in the reference frame $R$,
during a given time interval, also evaluated in that reference
frame, \emph{is the same in any two opposite directions.}
\par\smallskip

In a schematic $2$-dimensional Space-Time (or in a plane section
containing $\cal A$ of the \lq\lq true\rq\rq\ $4$-dimensional
Space-Time), the direction of isochronous subspaces is easily
obtained as shown on the left part of Figure~3: we take the two
light lines ${\cal L}^g$ and ${\cal L}^d$ through an event
$A\in{\cal A}$ (the red lines on that figure); we take another event
$A_1\in A$, for example in the future of $A$, and we build the
parallelogram $A\,A_1^g\,A_2\,A_1^d$ with two sides supported by
${\cal L}^g$ and ${\cal L}^d$, with $A$ as one of its apices and
$A_1$ as center. The isochronous subspaces are all the straight
lines parallel to the space-like diagonal $A_1^g\,A_1^d$ of that
parallelogram. Three of these lines are drawn (in blue) on Figure~3,
${\cal E}_{R,\,A}$, ${\cal E}_{R,\,A_1}$ and ${\cal E}_{R,\,A_2}$.
\par\smallskip

{---}\quad Why?
\par\smallskip

{---}\quad A light signal starting from $A$ covers, during the time
interval between events $A$ and $A_1$, the lengths $A_1\,A_1^g$
towards the left and $A_1\,A_1^d$ towards the right. These lengths
are equal because $A_1^g\,A_1^d$ is the diagonal of a parallelogram
whose center is $A_1$.

{---}\quad What for the \lq\lq true\rq\rq\ $4$-dimensional Minkowski
Space-Time $\cal M$~? And what are the Space ${\cal E}_R$ and the
Time ${\cal T}_R$ of our reference frame $R$?
\par\smallskip

\begin{figure}[H]
  $$\scalebox{1.2}{\includegraphics[
  ]{relpic.4}}$$
  \caption{Change of Galilean reference frame.}
\end{figure}
{---}\quad It is the same, as shown on the right side of Figure~3.
Take the event $A_2$ on the light line $\cal A$ such that $A_1$ is
the middle point of $A\,A_2$. Consider the future light half-cone
with apex $A$ and the past light half-cone with apex $A_2$. Their
intersection is a $2$-dimensional sphere $S$. The unique affine
hyperplane ${\cal E}_{R,\,A_1}$ which contains $S$ is an isochronous
subspace for the Galilean frame determined by the direction of $\cal
A$ (in blue on Figure~3). The other isochronous subspaces for that
Galilean frame are all the hyperplanes parallel to ${\cal
E}_{R,\,A_1}$. The Space ${\cal E}_R$ is the set of all the
isochorous lines, \emph{i.e} the set of all straight lines parallel
to $\cal A$, and the Time ${\cal T}_R$ the set of all isochronous
subspaces. Minkowski Space-Time $\cal M$ splits into the product
${\cal E}_R\times{\cal T}_R$, or in other words can be identified
with that product, because a pair made by an isochorous line and an
isochronous subspace determine a unique element of $\cal M$, the
event at which they meet.
\par\smallskip

{---}\quad What happens if you change your Galilean frame?
\par\smallskip

{---}\quad Of course, as for Galilean frames in Leibniz Space-Time,
the direction of isochorous lines (the straight world lines of
points at rest with respect to the chosen Galilean frame) is
changed. Moreover, contrary to what happened in Leibniz Space-Time,
the direction of isochronous subspaces is also changed! Therefore,
the chronological order of two events can be different when it is
appreciated in two different Galilean frames!

\subsection{Metric properties of Minkowski Space-Time}

Up to now, we have compared the lengths of two straight line
segments in $\cal M$ only when they were supported by parallel
straight lines. That was allowed by the \emph{affine structure} of
$\cal M$. We need more, because the spectral lines of atoms allow us
to build clocks and to compare time intervals measured in two
different Galilean frames.

\subsection{Comparison of times}
\begin{figure}[H]
  $$\scalebox{1.2}{\includegraphics[
  ]{relpic.5}}$$
  \caption{Comparison of times.}
\end{figure}

Let $A\,A_1$ and $A\,B_1$ be two straight line segments supported by
two different time-like straight lines $\cal A$ and $\cal B$, which
meet at the event $A$. Let $R_{\cal A}$ and $R_{\cal B}$ be the
Galilean frames determined by the directions of $\cal A$ and $\cal
B$, respectively. We assume that the time intervals corresponding to
$A\,A_1$ measured in $R_{\cal A}$, and to $A\,B_1$ measured in
$R_{\cal B}$, are the same. Let $B'$ be the event at which the
time-like straight line $\cal B$ meets the isochronous subspace
${\cal E}_{R_{\cal A},A_1}$ containing $A_1$ of the Galilean frame
$R_{\cal A}$ (figure~4). Since the events $A_1$ and $B'$ are
synchronous for $R_{\cal A}$, the time interval corresponding to
$A\,B_1$ appears longer than the time interval corresponding to
$A\,A_1$ when both are observed in the reference frame $R_{\cal A}$,
by the ratio $\displaystyle\frac{A\,B_1}{A\,B'}$. That ratio is the
\emph{ratio of dilation of times} of the Galilean frame of $R_{\cal
B}$, when observed in the Galilean frame $R_{\cal A}$. Similarly,
$\displaystyle\frac{A\,A_1}{A\,A'}$ is the ratio of dilation of
times of the Galilean frame $R_{\cal A}$ when observed in the
Galilean frame $R_{\cal B}$. According to the Principle of
Relativity, each one of these two Galilean frames must play the same role with
respect to the other, which implies the equality $\displaystyle
\frac{A\,A_1}{A\,A'}=\frac{A\,B_1}{A\,B'}$. By a well known property
of hyperbolae, that equality holds \emph{if and only if $A_1$ and
$B_1$ lie on the same arc of hyperbola which has the light lines
${\cal L}^d$ and ${\cal L}^g$ (which meet at $A$ and are contained
in the two-dimensional plane which contains $\cal A$ and $\cal B$)
as asymptotes}. Or more generally, on the same hyperboloid with the
light cone of $A$ as asymptotic cone.
\par\smallskip

\subsection{Comparison of lengths}

The comparison of lengths on two non-parallel space-like straight
lines is similar to the comparison of time intervals. Let $A\,A^d$
and $A\,B^d$ be two segments supported by two space-like straight
lines which meet at the event $A$. They are of equal length \emph{if
and only if $A^d$ and $B^d$ lie on the same hyperboloid with the
light cone of $A$ as asymptotic cone}.
\par\smallskip

\begin{figure}
  $$\scalebox{1.2}{\includegraphics[
  ]{relpic.6}}$$
  \caption{Comparison of lengths.}
\end{figure}

\section{Conclusion}

The comparison of time intervals and lengths presented above allows
a very natural introduction of the pseudo-Euclidean metric of
Minkowski Space-Time. The construction of isochronous subspaces in
two different Galilean frames, as presented above, leads to the
formulas for Lorentz transformations with a minimum of calculations.
The pictures we have presented allow a very easy explanation of the
apparent contraction of lengths and dilation of times associated to
a change of Galilean frames and a very simple explanation, without
complicated calculations, of the (improperly called) paradox of
Langevin's twins.
\par\smallskip

By explaining that the affine structure of Space-Time should be
questioned, a smooth transition towards General Relativity, suitable
from children from 8 to 108 years old, seems possible.
\par\smallskip\noindent
{\bf Acknowledgements.} The author thanks the team \lq\lq Analyse
alg\'{e}brique\rq\rq\ of the \lq\lq Institut de Math\'{e}matiques de
Jussieu\rq\rq\ and his University for taking in charge his
registration fee at this International Conference.


\begin{thebibliography}{9}

\bibitem{principles} Einstein, A., Lorentz, H.A., Weyl, H.,
Minkowski, H., {\it The Principles of Relativity\/}, a collection of
original papers on the special and general theory of relativity,
with notes by A.~Sommerfeld. Methuen and Company, 1923. Reprinted by
Dover Publications, Inc., New York.

\bibitem{newton} Newton, Isaac, {\it Principes
math\'{e}matiques de la Philosophie naturelle\/}, tomes I et II,
translated by Madame la Marquise du Chastellet, chez Desaint et
Saillant, Paris, 1759. Reprinted by the \'Editions Jacques Gabay,
Paris, 1990.

\bibitem{poinca} Poincar\'{e}, Henri,\quad {\it La M\'{e}canique nouvelle\/},
book containing the  text of a lecture presented at the congress of
the \lq\lq Association fran\c{c}aise pour l'avancement des
sciences\rq\rq\ (Lille, 1909), the paper dated 23 July 1905 {\it Sur
la dynamique de l'\'{e}lectron\/}, Rendiconti del Circolo matematico di
Palermo {\bf XXI} (1906), and a \lq\lq Note aux Comptes Rendus de
l'Acad\'{e}mie des Sciences\rq\rq\ with the same title dated 15 June
1905; Gauthier-Villars, Paris, 1924; reprinted by the \'Editions
Jacques Gabay, Paris, 1989.

\end{thebibliography}
\end{document}